\input harvmac
\input epsf.tex

\overfullrule=0mm
\def\file#1{#1}
\def\figbox#1#2{\epsfxsize=#1\vcenter{
\epsfbox{\file{#2}}}}
\newcount\figno
\figno=0
\def\fig#1#2#3{
\par\begingroup\parindent=0pt\leftskip=1cm\rightskip=1cm\parindent=0pt
\baselineskip=11pt
\global\advance\figno by 1
\midinsert
\epsfxsize=#3
\centerline{\epsfbox{#2}}
\vskip 12pt
{\bf Fig. \the\figno:} #1\par
\endinsert\endgroup\par
}
\def\figlabel#1{\xdef#1{\the\figno}}
\def\encadremath#1{\vbox{\hrule\hbox{\vrule\kern8pt\vbox{\kern8pt
\hbox{$\displaystyle #1$}\kern8pt}
\kern8pt\vrule}\hrule}}
\def\omit#1{}

\def\pre#1{({\tt
#1})}

\def\IR{\relax{\rm I\kern-.18em R}}


\nref\BIBLE{D. Bressoud, {\it Proofs and confirmations. The story of the alternating
sign matrix conjecture}, Cambridge University Press (1999).}
\nref\RS{A.V. Razumov and Yu.G. Stroganov, 
{\it Combinatorial nature
of ground state vector of O(1) loop model}, preprint \pre{math.CO/0104216},
Theor. Math. Phys. {\bf 138} (2004) 333-337.}
\nref\MNosc{S. Mitra and B. Nienhuis, {\it 
Osculating random walks on cylinders}, in
{\it Discrete random walks}, 
DRW'03, C. Banderier and
C. Krattenthaler edrs, Discrete Mathematics and Computer Science
Proceedings AC (2003) 259-264, \pre{math-ph/0312036} .} 
\nref\MNdGB{S. Mitra, B. Nienhuis, J. de Gier and M.T. Batchelor,
{\it Exact expressions for correlations in the ground state 
of the dense $O(1)$ loop model}, preprint \pre{cond-math/0401245}}
\nref\Wie{B. Wieland, {\it  A large dihedral symmetry of the set of
alternating-sign matrices}, 
Electron. J. Combin. {\bf 7} (2000) R37, 
\pre{math.CO/0006234}.}
\nref\LGV{B. Lindstr\"om, {\it On the vector representations of
induced matroids}, Bull. London Math. Soc. {\bf 5} (1973)
85-90\semi
I. M. Gessel and X. Viennot, {\it Binomial determinants, paths and
hook formulae}, Adv. Math. { \bf 58} (1985) 300-321. }

%

\Title{SPhT-T04/100}
{\vbox{
\centerline{A refined Razumov-Stroganov conjecture}
}}
\bigskip\bigskip
\centerline{P.~Di~Francesco,} 
\medskip
\centerline{\it  Service de Physique Th\'eorique de Saclay,}
\centerline{\it CEA/DSM/SPhT, URA 2306 du CNRS,}
\centerline{\it F-91191 Gif sur Yvette Cedex, France}
\bigskip
\bigskip\noindent
We extend the Razumov-Stroganov conjecture relating the groundstate
of the O(1) spin chain to alternating sign matrices, by relating
the groundstate of the monodromy matrix of the O(1) model
to the so-called refined alternating sign matrices, i.e.
with prescribed configuration of their first row, as well as to
refined fully-packed loop configurations on a square grid,
keeping track both of the loop connectivity and of the configuration
of their top row. We also conjecture
a direct relation between this groundstate and refined totally 
symmetric self-complementary plane partitions, namely, 
in their formulation as sets of non-intersecting 
lattice paths, with prescribed last steps of all paths.


\Date{07/2004}

%
%

\newsec{Introduction and basic definitions}

The proof of the alternating sign matrix conjecture is one of the keystones
of today's combinatorial world, borrowing ideas and concepts from both 
physics and mathematics (see Bressoud's beautiful book \BIBLE\ for
a complete history and references). An alternating sign matrix (ASM)
is a matrix of size $n\times n$ with entries $0,1,-1$ such that 
along each row and column, read in any direction, 
$+1$'s and $-1$ alternate, possibly separated by 
arbitrary numbers of $0$'s, and always starting with a number of zeros followed by $+1$. 
Note that the top row of such a matrix
contains exactly one $1$ and $n-1$ zeros. Accordingly, the number
$A_{n,j}$ of ASM of size $n\times n$ with a 
$1$ on top of the $j$th column reads
\eqn\asmref{ A_{n,j}= {n+j-2\choose j-1} {(2n-j-1)!\over (n-j)!} \prod_{i=0}^{n-1}
{(3i+1)!\over (n+i)!} }
Note also that $A_n\equiv A_{n+1,1}=\sum_{j=1}^n A_{n,j}=\prod_{0\leq i\leq n-1}(3i+1)!/(n+i)!$ 
is the total number of ASM of size $n\times n$.

Alternating sign matrices may alternatively be viewed as configurations
of the Fully-Packed Loop (FPL) model with special boundary
conditions on an $n\times n$ square grid. In this model, edges of the square lattice
are occupied or not by bonds in such a way that exactly two bonds are adjacent
to every vertex of the lattice. This gives rise to six possible vertex configurations,
two of which are ``crossings" (opposite bonds), and the rest of which are ``turns" 
(bonds at a right angle). On the square $n\times n$ grid, we impose the boundary condition
that every other outer edge perpendicular and adjacent to the boundary is occupied by a bond.
The configurations of such a model are in bijection with ASM of size $n\times n$. 
This is best seen by applying the following dictionary
between $1,0,-1$ ASM entries and FPL vertex configurations on even or odd vertices (according
to the parity of the sum of their coordinates on the grid):
\eqn\dictio{\figbox{10.cm}{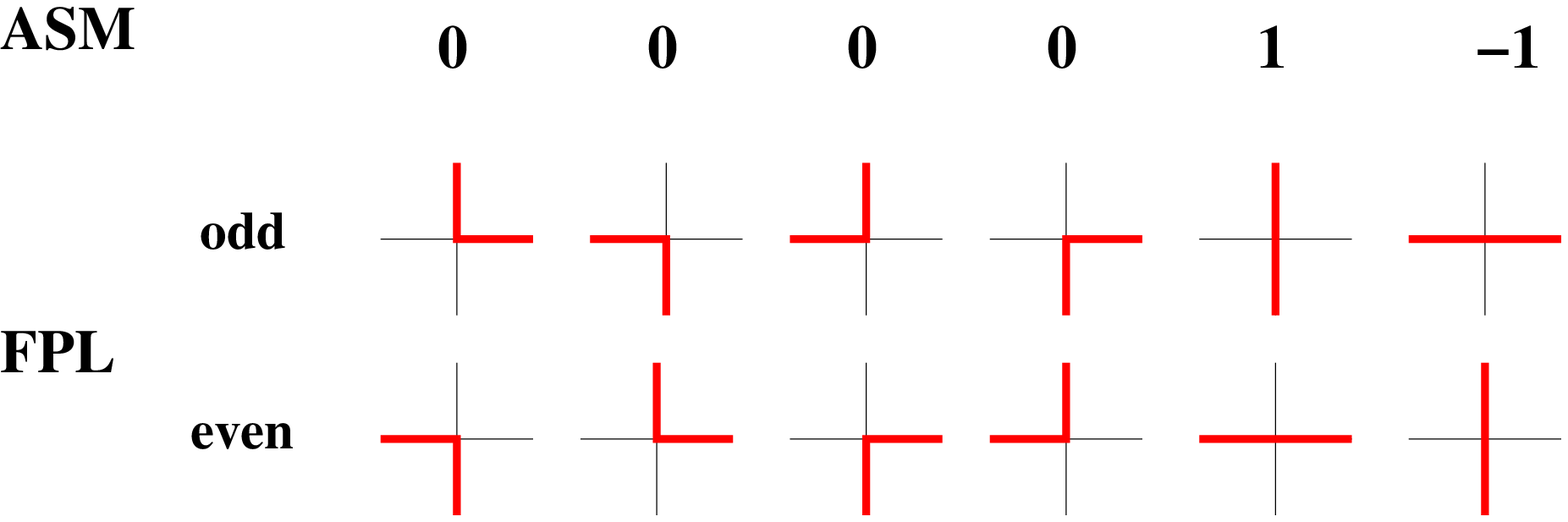} }
As a consequence, $A_{n,j}$ also counts the FPL configurations with a ``crossing" vertex
at the top of the $j$th column. 
\fig{The $17$ FPL configurations corresponding to the connectivity
$(21)(54)(76)(98)(10\, 3)$, at $n=5$. We have represented the edges occupied by bonds
in thick red lines. The complete grid, together with the labeling of
external bonds, is represented only on the top left configuration.}{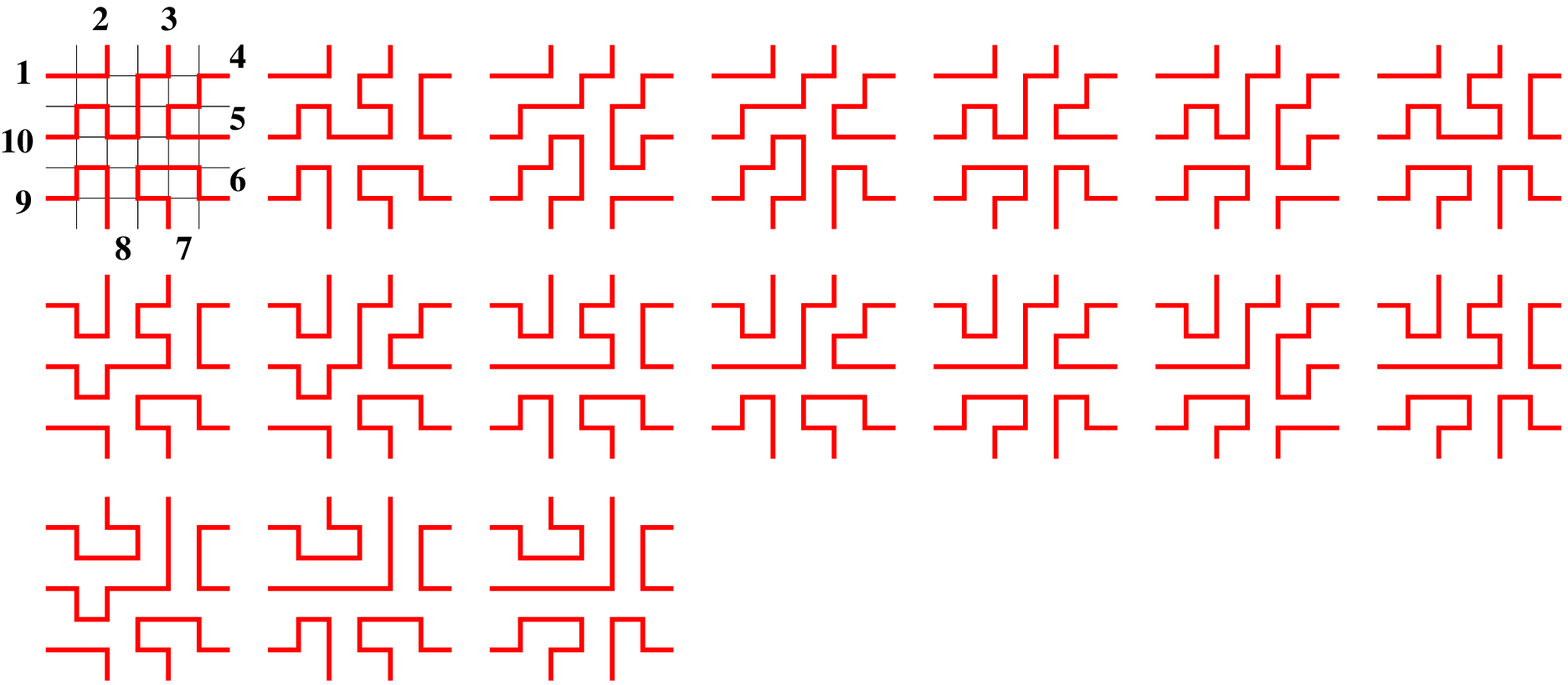}{13.cm}
\figlabel\seventeen

Notice that the bonds of the FPL form non-intersecting broken lines and loops on the grid.
In particular, we may keep track of the connectivity of the external bonds, say labeled
$1$ to $2n$, starting from the top left in clockwise direction around the boundary, in the form
of the list of pairs of connected points (by convention and for later convenience, we decide
to write first the larger of the two labels in each pair).
This is illustrated in Fig.\seventeen, where we have represented for $n=5$ the $17$
FPL configurations corresponding to the connectivity $(21)(54)(76)(98)(10\, 3)$.
Such a connectivity is also called a {\it link pattern}, sometimes represented
as a chord diagram
in which $2n$ labeled points around a circle are connected by pairs via 
non-intersecting segments (chords). We denote by $LP_n$ the set of link patterns
on $2n$ points, with cardinality $|LP_n|=c_n=(2n)!/(n!(n+1)!)$, the $n$th Catalan number. 
For $\pi\in LP_n$, we denote by $A_n(\pi)$ the total number of FPLs whose external bonds
are connected according to the link pattern $\pi$. The dihedral group $D_{2n}$ acts 
on link patterns via its two generators $r$ and $s$,
namely the rotation $r$ by one unit, i.e. the
relabelling of all points $i\to i-1$ for $i=2,...,2n$, and $1\to 2n$, and the basic reflection
$s$, i.e. the relabelling of all points $i\to 2n+2-i$, with the two fixed points $1$ and $n+1$.
In \Wie, Wieland showed how to implement this action on FPL configurations, with 
the consequence that $A_n(r\pi)=A_n(s\pi)=A_n(\pi)$ for all link patterns $\pi$.

On an apparently completely unrelated front, 
Razumov and Stroganov \RS\ have studied the groundstate of the spin chain underlying the O(1)
lattice model, and found a remarkable identity between the $A_n(\pi)$ and the entries of
the groundstate eigenvector of the chain. In its simplest formulation, the O(1) spin chain's
Hamiltonian is the sum of the generators $e_i$ of the Temperley-Lieb algebra with cyclic boundary conditions,
$CTL_{2n}(1)$. Going to the link pattern representation,
these generators act on link patterns $\pi\in LP_n$ as follows. Assume
the points $i$ and $i+1$ are connected to respectively points $k$ and $m$ in $\pi$. Then 
$e_i|\pi\rangle=|\pi'\rangle$
is the link pattern in which the pairs $(i+1,i)$ and $(k,m)$ are connected, while all other pairs remain
unaltered. In particular,
if $k=i+1$ and $l=i$, then $e_i\pi=\pi$. When $i=2n$, we must set the point $2n+1\equiv 1$ (cyclic
boundary conditions). This gives rise to a $c_n\times c_n$ matrix representation for each $e_i$.
Note that $e_i$ sends each link pattern $\pi$ to a single link pattern $\pi'$, therefore
it has the obvious left eigenvector $v_n=(1,1,...,1)$ with eigenvalue $1$, namely $v_n e_i=v_n$
for all $i$. Moreover the matrix representing $e_i$ only has zeros and ones.
The $e_i$'s satisfy the defining relations of $CTL_{2n}(1)$, namely that
$e_i^2=e_i$, $[e_i,e_j]=0$ as soon as $|i-j|>1$, and $e_ie_{i\pm 1}e_i=e_i$, with the convention 
that $e_{2n+1}\equiv e_1$.
Let $H=\sum_{i=1}^{2n} e_i$ be the Hamiltonian of
the O(1) spin chain with cyclic boundary conditions, and $\Psi_n$ its ``groundstate" vector
(here with largest eigenvalue),
\eqn\eigen{\Psi_n=\sum_{\pi\in LP_n} B_n(\pi) |\pi\rangle} 
As $H$ is represented by a $c_n\times c_n$
matrix with non-negative integer entries, the largest eigenvalue corresponds to the Perron-Frobenius
eigenvector with positive entries. But $v_n$ is obviously a left eigenvector of $H$ for the eigenvalue
$2n$, as $v_n H=v_n\sum e_i=2n v_n$, and as its entries are positive it is the left Perron-Frobenius 
eigenvector, and $\Psi_n$ is the right Perron-Frobenius eigenvector, with same eigenvalue $2n$,
hence 
\eqn\frope{H\Psi_n=2n\Psi_n}
The entries $B_n(\pi)$ are positive and may be chosen to be integers.
The Razumov-Stroganov conjecture \RS\ (still a conjecture to this day)
simply states that one choice leads to 
\eqn\rastro{ B_n(\pi)=A_n(\pi), \ \ {\rm for} \ \ {\rm all}\ \ \pi\in LP_n}
This has been extended to a number of symmetry classes of ASM, upon modifying boundary
conditions in the O(1) spin chain (see for instance Refs. \MNosc\ \MNdGB).

We see that two types of very different observables have been defined on FPLs: (a) a local one, 
keeping track of the crossing vertices at the top boundary (and leading to the numbers $A_{n,j}$),
and (b) a very non-local one, keeping track of the connectivity of the $2n$ boundary bonds.
In this note, our aim is to extend the Razumov-Stroganov conjecture so as to include the
observables (a) as well. To this end, for $\pi\in  LP_n$ and 
$1\leq j\leq n$, we introduce ``refined FPL numbers" $A_{n,j}(\pi)$
which count the total number of ASM with a $1$ on top of their $j$th column, and
such that the corresponding FPL has link pattern $\pi$. For instance, the three lines of Fig.\seventeen\
correspond to respectively the $7$, $7$, and $3$ FPL configurations at $n=5$ with link pattern 
$\pi=(21)(54)(76)(98)(10\, 3)$ and corresponding to $j=1$, $2$ and $4$ respectively (recall $j$
is the position of the crossing vertex along the top line of the configuration,
counted here from $1$ to $5$): in this case, we have
$A_{5,1}(\pi)=A_{5,2}(\pi)=7$, $A_{5,3}(\pi)=A_{5,5}(\pi)=0$ and $A_{5,4}(\pi)=3$.
In appendix A, we list the first few of these numbers for up to $n=5$, in the form of 
``refined FPL vectors" $\alpha_n(t)$ with entries 
\eqn\entryalpha{ \alpha_n(\pi|t)=\sum_{j=1}^n A_{n,j}(\pi) t^{j-1} }
listed in increasing lexicographic order on the link patterns expressed as lists of pairs of
connected points as above.
Our main conjecture will
connect these numbers to some object defined purely in terms of the Cyclic Temperley-Lieb
algebra generators $e_i$.

Let us make clear that this note is purely exploratory and contains
no proof, and that the main result is simply a conjecture. This conjecture however
sheds new light on the Razumov-Stroganov conjecture, and hints at a possible direction for proving it.

We also comment
on the relation to Totally Symmetric Self-Complementary Plane Partitions (TSSCPP), in the form
of Non-Intersecting Lattice Path (NILP) counting, and propose
a refinement in the same spirit as some conjectures of \MNosc\ \MNdGB, which may possibly 
set the grounds for a bijective proof of the ASM-TSSCPP correspondence.

This note is completed by appendices A through E, where various numbers are gathered up to size
$n=5$. This limit on size is chosen somewhat arbitrarily (say in order for lists to fit in a page), as
today's computers allow to reach larger numbers. 

\newsec{The refined Razumov-Stroganov conjecture}

\subsec{O(1) model's monodromy matrix}

The transfer matrix of the O(1) model, as particular case of the six vertex model,
say on a cylinder of square lattice of even width $2n$, may be characterized via 
a standard solution of the Yang-Baxter equation
\eqn\ybe{ X_i(v) X_{i+1}(v+w) X_i(w)=X_{i+1}(w)X_i(v+w)X_{i+1}(v) }
where the ``face transfer matrix" operator $X_i(u)$ acts on a row by locally adding at site $i$ 
a Boltzmann weight of the model.
The solution involved in the definition of the O(1) model reads
\eqn\soly{ X_i(v)={I}+ f(v) e_i, \qquad f(v)={e^v-1\over q-q^{-1}e^v}, \quad q+1/q=1 }
where $e_i$ is a generator of the Cyclic Temperley-Lieb algebra $CTL_{2n}(1)$, acting on a (cyclic) row
of size $2n$, and represented via a $c_n\times c_n$ matrix as above. 
For later convenience, we prefer to use the following definition for $X_i$:
\eqn\betterx{ X_i(t)\equiv  t {I} +(1-t) e_i }
for some real parameter $t$. 

Using the $X_i(t)$, one may introduce various transfer matrices for the model, upon taking a suitable
product of them, and a trace over the ``auxiliary space".
Here we are more interested in the {\it monodromy matrix} of the model, defined as 
\eqn\mono{ M_n(t)=X_1(t)X_2(t)\cdots X_{2n}(t) }
Note that, as opposed to transfer matrices, monodromy matrices do not commute at different 
values of $t$, this is precisely what makes them interesting from our point of view.

Let us perform an expansion of $M_n(t)$ around $t=1$:
\eqn\expaam{ M_n(1-\epsilon)=\prod_{i=1}^{2n}({I}+\epsilon(e_i-I))=I+\epsilon H_0
+{\epsilon^2\over 2} (H_0^2-H_0+H_1+[e_1,e_{2n}])+O(\epsilon^3)}
where we have used the notation
\eqn\notaH{\eqalign{ H_0&=H-2n I=\sum_{i=1}^{2n} (e_i-I) \cr
H_1&=\sum_{i=1}^{2n} [e_i,e_{i+1}]\cr}}
with the convention that $e_{2n+1}=e_1$.
Here $H_0$ denotes the Hamiltonian of the O(1) spin chain (including an unimportant shift
by a constant diagonal), and $H_1$ its first conserved quantity, namely such that $[H_0,H_1]=0$.
The expansion \expaam\ shows the occurrence of an extra term $[e_1,e_{2n}]$ that {\it does not commute}
with $H_0$ and $H_1$. This explains why two such monodromy matrices taken at distinct values of $t$
do not commute.

\subsec{The groundstate vector of the monodromy matrix}

As before, and taking $0<t<1$, we see that each $X_i(t)$ is a linear combination of two
$c_n\times c_n$ matrices with non-negative integer entries, hence we may apply the Perron-Frobenius theorem
to their product $M_n(t)$.
On the other hand, the left eigenvector $v_n$ defined above is still a left eigenvector
for all $X_i(t)$, with eigenvalue $1$, hence $v_nM_n(t)=v_n$ as well. Let $\Psi_n(t)$ denote the
right Perron-Frobenius eigenvector of $M_n(t)$, then we have
\eqn\pfro{ M_n(t)\ \Psi_n(t)= \Psi_n(t) }

Using the expansion \expaam\ to all orders in $\epsilon$, we may formally expand 
$\Psi_n(1-\epsilon)=\sum_{m\geq 0} \epsilon^m \Psi_n^{(m)}$, and express \pfro\
order by order in $\epsilon$. At order zero, we get a tautology, while at order $1$, we 
get 
\eqn\ordone{ H_0 \Psi_n^{(0)}=0 }
This is nothing but the eigenvalue equation \frope, and we may normalize $\Psi_n(t)$ in
such a way that $\Psi_n^{(0)}=\Psi_n(1)=\Psi_n$ of Razumov-Stroganov \eigen, suitably normalized 
for its entries to match the $A_n(\pi)$.
Now eq. \pfro\ provides us with a pertubative series for $\Psi_n(t)$ in the vicinity of $t=1$.

Let $\Psi_n(\pi|t)$ denote the entry of $\Psi_n(t)$ corresponding
to the link pattern $\pi\in LP_n$. The Razumov-Stroganov conjecture amounts to 
$\Psi_n(\pi|1)=A_n(\pi)$.
The first few vectors $\Psi_n(t)$ up to $n=5$ are listed in appendix B below, 
with  entries listed in increasing lexicographic
order on the sequence of pairs of connected labels, like in the case of the vector $\alpha_n(t)$
of appendix A.

\subsec{Claims and main conjecture}

We now present a series of claims regarding the vector $\Psi_n(t)$, all of which are satisfied
by the data of appendix B. 

\noindent{\bf Claim 1:} $\Psi_n(t)$ is a polynomial of degree $n-1$ of $t$, with only
non-negative integer coefficient entries.

\noindent{\bf Claim 2:} $\Psi_n(0)=\{\Psi_{n-1}(1),0,0,\cdots, 0\}$, where the last vector is completed
by $c_n-c_{n-1}$ zeros.

\noindent{\bf Claim 3:} We have the reflection relation 
\eqn\weha{ \Psi_n(\pi|t)=t^{n-1} \Psi_n(s\pi|{1\over t}) }
Note that this reduces to the Wieland invariance of the FPL numbers
under the reflection $s$ at $t=1$, modulo the Razumov-Stroganov conjecture. 

The behavior under rotations is less clear, 
and coefficients get reshuffled, although their sum remains constant. 

Let us now compare the vectors $\alpha_n(t)$ and $\Psi_n(t)$.
Comparing the data in appendices A and B,
we observe that in general $\Psi_n(\pi|t)\neq A_n(\pi|t)$.
In view of the above bad behavior under rotations of the link patterns, this suggests 
to consider instead the sums of entries of $\Psi_n(t)$ over the orbits of link patterns
under the rotation $r$. If $\ell_n(\pi)$ denotes the length of the orbit of the link pattern
$\pi$ under $r$, namely the smallest positive integer $\ell$, such that $r^\ell\pi=\pi$, 
we define the polynomials
\eqn\defnum{ \gamma_n(\pi|t)=\sum_{\ell=0}^{\ell(\pi)-1} \Psi_n(r^\ell\pi|t) =\sum_{j=1}^n
\gamma_{n,j}(\pi) t^{j-1} }
We now come to the main conjecture of this paper:

\noindent{\bf Conjecture 1:} The numbers $\gamma_{n,j}(\pi)$ count the total number
of alternating sign matrices with a $1$ at the top of their $j$th column {\it and}
such that in the FPL picture, {\it up to a rotation on the labels}, the connectivity of 
their external bonds is given by the link pattern $\pi$. 
In other words, we have
\eqn\mainconj{ \gamma_{n,j}(\pi)=\sum_{\ell=0}^{\ell(\pi)-1} A_{n,j}(r^\ell\pi) }
As a corollary, the sum of
all the entries in $\Psi_n(t)$ is nothing but
\eqn\sumofentries{ v_n\cdot \Psi_n(t)=\sum_{\pi\in LP_n}\Psi_n(\pi|t)=
\sum_{j=1}^n t^{n-1} A_{n,j}}
namely the generating polynomial for the refined ASM numbers $A_{n,j}$.

We display the first few polynomials $\gamma_n(\pi|t)$ for $n$ up to $5$ in appendix C below.

\noindent{\bf Claim 4:} The first entry $\Psi_n(\pi_0\equiv (21)(43)...(2n\, 2n-1)\|t)$ of $\Psi_n(t)$
is the sum of all entries of $\Psi_{n-1}(t)$,
namely
\eqn\sumentries{ \Psi_n(\pi_0|t)= v_{n-1}\cdot \Psi_{n-1}(t)= \sum_{\pi\in LP_{n-1}}\Psi_{n-1}(\pi|t)}
This reduces at $t=1$ to another conjecture by Razumov and Stroganov. Using conjecture 1 above, 
we see that the $A_{n-1,j}$ are entirely coded in this first entry.

\newsec{A connection between TSSCPP and the vector $\Psi_n(t)$.}

\subsec{TSSCPP via NILP}

\fig{A sample TSSCPP at $2n=12$ (left). We have extracted a fundamental domain
and represented the corresponding rhombus tiling of $1/12$ of the
original hexagon (center). Marking the chains of pink and yellow tiles with red lines
leads to NILP (right) going from the $n-1=5$ green points to the black ones with right up
or down steps only, taken along the edges of the underlying rhombic lattice (dashed lines). 
We have represented by empty black circles the $n-1=5$ remaining
attainable endpoints.}{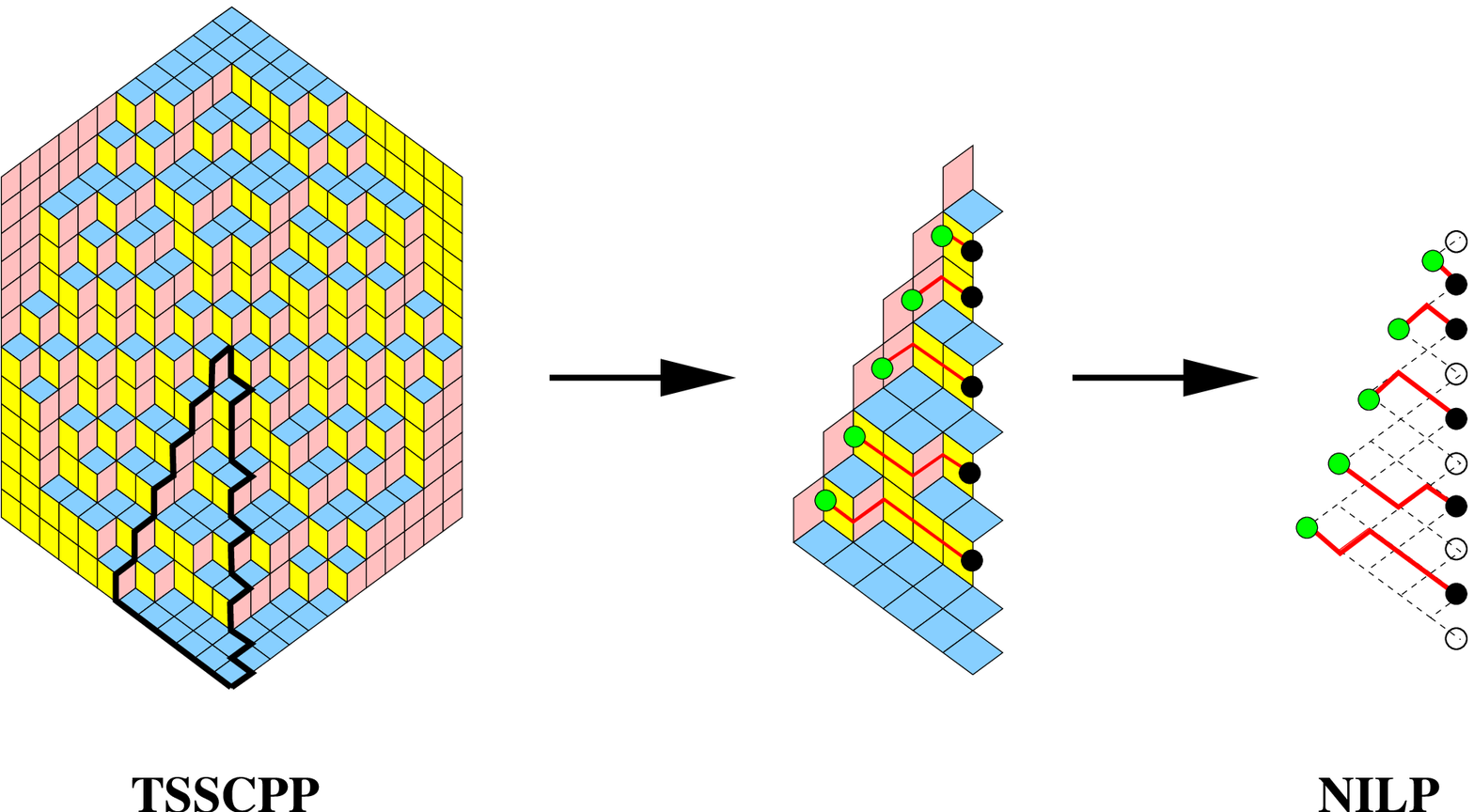}{13.cm}
\figlabel\tsscpp

Plane partitions are piling-ups of unit cubes inside some rectangle parallelepipedon of $\IR^3$,
with sides of integer lengths $a,b,c$ and corner at the origin, and with the constraint that 
``gravity" has direction $-(1,1,1)$, cubes are therefore intuitively attracted to the origin corner, 
and occupy successive positions on the cubic lattice. When viewed in perspective from the
$(1,1,1)$ direction, such piling-ups form rhombus tilings of an hexagon with sides of lengths
$a\times b\times c$, by means of the three unit rombi obtained by gluing pairs 
of equilateral triangles along an edge in the triangular lattice 
(the rhombi are the tops of the piled-up cubes seen in perspective).
Here we concentrate on the so-called
Totally Symmetric Self-Complementary Plane Partitions (TSSCPP), for which $a=b=c=2n$, and which
are identical to their complement in the cube of size $2n$ and 
invariant under all reflections of the regular hexagon of size $2n$ in the tiling picture 
(see Fig.\tsscpp\ for an example with $2n=12$).  
One remarkable property is that the total number of TSSCPP is nothing but the ASM number
$A_n$. The proof of this fact is reviewed in \BIBLE, with a wealth of material and
references on plane partitions. Let us point out that this proof is all but bijective, and 
finding a direct bijection between TSSCP and ASM remains an open challenge.

As described in \BIBLE,
TSSCP however
are known to be in bijection
with special sets of non-intersecting lattice paths (NILP), which roughly speaking follow
in the rhombus tiling picture the
sequences formed by two of the three types of rhombi, forming non-intersecting 
chains across a fundamental domain
equal to $1/12$ of the total hexagon (see Fig.\tsscpp\ for an illustration). 
Upon an unimportant deformation, the latter\foot{We now slightly change
the geometry of these paths, compared to those defined in \BIBLE, for the sake 
of simplicity and to best suit our needs here.} are sets of $n-1$
non-intersecting oriented paths joining vertices
of the integer square lattice, with only ``vertical" and ``diagonal"
steps of length $1$ or $\sqrt{2}$, respectively
joining points $(i,j)\to (i,j+1)$ and $(i,j)\to (i+1,j+1)$. 
Moreover, they must start at points $(j,-j)$, $j=1,2,...,n-1$ and end up at points
$(j,0)$, $j=1,2,...,n-1$. The total number of such paths is nothing but that
of alternating sign matrices $A_n$.

\fig{The FPL configurations for $n=3$, arranged according to their
corresponding link patterns, expressed as lexicographically ordered sequences
of pairs of connected labels, the largest labels always coming first
in the pairs. We have indicated by circles
the position of the ``crossing" vertex in the first row of each FPL configuration.
The link patterns
are represented as arch configurations, or equivalently as Dyck paths. 
Recording the position of up-steps (excluding the first) yields
an admissible pair of integers $\{r_1,r_2\}$. We have represented the
NILP with endpoints corresponding to these pairs.}{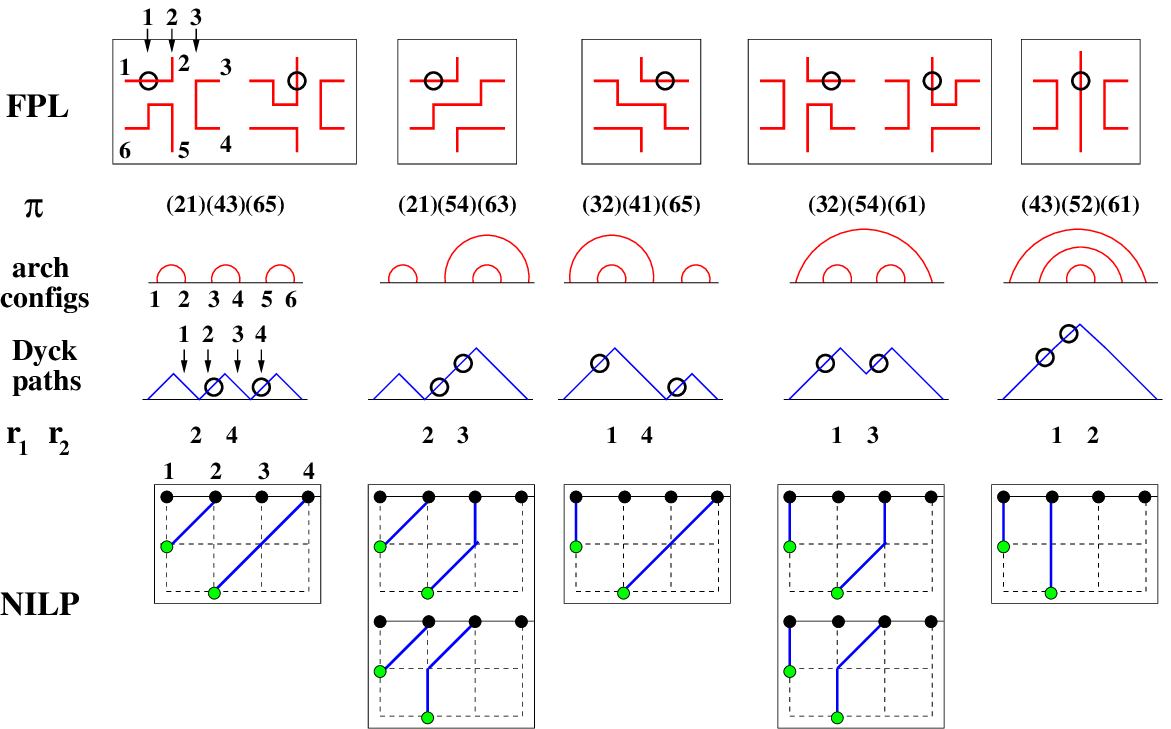}{13.cm}
\figlabel\fplnilp

Let us compute the total number of NILP with prescribed ``arrival points" say $(r_i,0)$,
with $1\leq r_1<r_2<...<r_{n-1}\leq 2(n-1)$. In fact, it is easy to convince oneself
that the $j$th path, starting at $(j,-j)$ must end at the $j$th endpoint along the x axis,
namely $(r_j,0)$. Moreover, to reach the farthest point to the right, one must only make diagonal
steps, hence the constraint that $r_j\leq 2j$ for all $j$. Let $P_n(r_1,r_2,...,r_{n-1})$
denote the total number of NILP with endpoints $(r_j,0)$, $j=1,2,...,2n-2$.
According to the so-called Lindstr\"om-Gessel-Viennot (LGV) formula \LGV, this number is easily expressed
as a determinant. Let $P_{i,r}$ denote the total number of paths, with vertical
and right diagonal up-steps only, joining $(i,-i)$ to $(r,0)$, then
\eqn\ezp{ P_{i,r}={i\choose r-i} }
as we must pick $r-i$ diagonal steps among a total of $i$. The LGV formula allows to express
\eqn\allow{ P_n(r_1,r_2,...,r_{n-1})=\det\left(P_{i,r_j}\right)_{1\leq i,j\leq n-1} }
These are nothing but the various non-vanishing minors of the $(n-1)\times (2n-2)$ matrix
with entries \ezp. Note that these minors are non-vanishing for strictly
ordered $(n-1)$-tuples of $r$'s, iff $r_j\leq 2j$ for all $j$, and $r_1\geq 1$.
The corresponding $(n-1)$-tuples will be called {\it admissible}.
Now comes the fundamental remark that
there are exactly $c_n$ admissible $(n-1)$-tuples, $c_n$ the $n$th Catalan number, and moreover
there is a natural bijection between these and the link patterns of $LP_n$. To get it,
we simply list the admissible $(n-1)$-tuples in {\it decreasing} lexicographic order, and match
them with the link patterns, listed as before in increasing lexicographic order.
More directly, let us draw the link patterns as $2n$ points on a line, connected via
non-intersecting semi-circles in the upper half-plane, as shown in Fig.\fplnilp\ for the case $n=3$.
To each such ``arch configuration", we associate a so-called Dyck path, namely a path with a
total of $n$ up and $n$ down unit steps, that stays above its origin, with the rule that,
visiting the arch configuration from left to right, we have a step up if an arch starts,
and a step down if an arch terminates. Now the increasing sequence $r_1,..,r_{n-1}$ is just 
that of the positions
of the up steps along the Dyck path, excluding the first one (which is always up, at position $0$), 
counted from $1$ to $2n-2$ (the last position $2n-1$ always corresponds to a down step).
This correspondence allows to associate to any link pattern a set of NILP with prescribed 
starting and endpoints.

\subsec{The refined TSSCPP vectors}

The numbers $P_n(r_1,r_2,...,r_{n-1})$ may be alternatively computed by induction on $n$, by
noticing that if the paths are stopped one step before their ends, they still all end up
on $n-1$ points of a horizontal line (with y coordinate $-1$), whose x coordinates 
differ from the $r$'s by $0$ or $-1$.
This turns into the recursion relation
\eqn\recuP{ P_n(r_1,r_2,...,r_{n-1})=\sum_{t_1,t_2,..,t_{n-1}\in \{0,1\}\atop
1\leq r_1-t_1<r_2-t_2<...<r_{n-1}-t_{n-1} }
P_{n-1}(r_2-t_2,r_3-t_3,...,r_{n-1}-t_{n-1}) }
where $P_n$ is understood to vanish whenever its argument is not admissible.
This allows for defining refined TSSCPP numbers, according to the number of vertical last steps
of the corresponding NILP. Letting $P_{n,m}(r_1,...,r_{n-1})$ denote the total number of NILP
ending up at $(r_j,0)$, $j=1,2,...,n-1$ and with exactly $m$ final vertical steps,
we introduce the polynomial
\eqn\intropo{ P_n(r_1,...,r_{n-1}|t)=\sum_{m=0}^{n-1} t^m P_{n,m}(r_1,...,r_{n-1}) }
The latter is easily derived by slightly altering the recursion relation \recuP:
\eqn\recupt{ P_n(r_1,...,r_{n-1}|t)=\sum_{t_1,t_2,..,t_{n-1}\in \{0,1\}\atop
1\leq r_1-t_1<r_2-t_2<...<r_{n-1}-t_{n-1} } t^{\Sigma_{1\leq i\leq n-1} t_i}
P_{n-1}(r_2-t_2,r_3-t_3,...,r_{n-1}-t_{n-1})}
By a slight abuse of notation, we denote by $P_n(\pi|t)\equiv P_n(r_1,r_2,...,r_{n-1}|t)$, upon
using the abovementioned bijection between link patterns and admissible $(n-1)$tuples, 
and build a vector $P_n(t)=\{ P_n(\pi|t) \}_{\pi\in LP_n}$.
We have listed the first few such ``refined TSSCPP vectors" in appendix D below.

We come to the main claim of this section:

\noindent{\bf Claim 5:} The sum of entries of $P_n(t)$ is nothing but 
$A_n(t)=\sum_{1\leq j\leq n-1}A_{n,j} t^{j-1}$.

This claim might be directly related to the so-called gog-magog correspondence used to prove 
the refined  ASM conjecture and referred to
in \BIBLE. It also suggests there might exist a direct link between the vectors $P_n(t)$ and
$\Psi_n(t)$.

\subsec{A further relation between entries of $P_n(t)$ and $\Psi_n(t)$}

Considering the entries of $\Psi_n(t)$ as relative probability weights for link patterns,
we may compute various expectation values. In particular, in the context of the O(1)
spin chain, Mitra et al. \MNosc\ \MNdGB\ considered the probability $p_n(m)$ that link patterns have 
the $m$ consecutive points $1,2,...,m$ connected to other points $j>m$. The lexicographic
order chosen here to list the entries of $\Psi_n(t)$ allows to write the corresponding
probability $p_n(m|t)$ in a very simple manner. We now denote by $\Psi_{n,k}(t)$,
$k=1,2,...,c_n$ the entries of the vector corresponding to the lexicographic order 
of link patterns. Then
\eqn\wripmn{ p_n(m|t)= {1\over v_n\cdot \Psi_n(t)} \sum_{k=B_{n,m}}^{c_n} \Psi_{n,k}(t) }
The numbers $B_{n,m}$ are related to the numbers $b_{n,m}$ of link patterns in which
points $1,2,...,m$ are connected to points with $j>m$ and $m$ is connected to $m+1$. 
Using the abovementioned correspondence with Dyck paths, $b_{n,m}$ is the number of Dyck paths
of $2n$ steps starting with $m$ consecutive up-steps, and one down-step, easily evaluated to be
\eqn\bnmnum{ b_{n,m}={2n-m-1\choose n-1}-{2n-m-1\choose n} }
for $1\leq m\leq n$.
Then as the lexicographic order simply lists the link patterns in increasing number $m$
of such consecutive points connected outside of their set, \wripmn\ follows, with
\eqn\Bmndef{B_{n,m}=1+\sum_{j=1}^{m-1} b_{n,j} }
for $m=1,2,...,n$.
In particular, we have $B_{n,1}=1$ so that $p_n(1)=1$. On the other extreme, we have
$B_{n,n}=c_n$, hence the sum in the numerator of \wripmn\ reduces to the last entry, with value $t^{n-1}$.

Performing analogous sums of entries of the refined TSSCPP vector $P_n(t)$, we have
come to the final conjecture of this note:

\noindent{\bf Conjecture 2:} The sums of entries from  $B_{n,m}$ to $c_n$ coincide for
all $m=1,2,...,n$ in $P_n(t)$ and in $\Psi_n(t)$, namely
\eqn\phidef{\varphi_n(m|t)\equiv \sum_{k=B_{n,m}}^{c_n} \Psi_{n,k}(t) =\sum_{k=B_{n,m}}^{c_n} 
P_{n,k}(t) }
where $P_{n,k}(t)$ denotes the $k$th entry of $P_n(t)$.

These sums of entries $\varphi_n(m|t)$ are listed up to $n=5$ in appendix E below.
Note that the last sum, reduced to the last entry of both vectors $\Psi_n(t)$ and $P_n(t)$, 
is $t^{n-1}$.
In NILP language, this is the contribution of the single
set of paths whose steps are all vertical.
At $t=1$, $\varphi_n(m|1)$ were conjectured in \MNosc\ \MNdGB\ to be given by simple product formulas.
Viewed as sums of entries in the vector $P_n(1)$ such formulas are much less mysterious,
as the entries of $P_n(1)$ are simply minors of a matrix with simple combinatorial entries
\allow.

\newsec{Conclusion and discussion}

In this note, we have presented two main conjectures regarding the entries 
of the Perron-Frobenius
eigenvector $\Psi_n(t)$ of the monodromy matrix $M_n(t)$ of the O(1) model, expressed 
purely in terms of the generators of the Cyclic Temperley-Lieb algebra.

The first one allows to interpret the sum of entries over the orbit of each link pattern under 
rotations as the generating function for ASM with a $1$ at a fixed position
in the top row and whose connectivity in the FPL language corresponds to some link pattern in the orbit.
Note here that we have had to resolve the fact that imposing a condition on the first row
breaks the cyclic invariance of the FPL counting problem: this is the reason why
we have to sum over all rotated link patterns in each orbit.
The fact that $\Psi_n(\pi|t)$ do not match individually the generating function $A_n(\pi|t)$
for ASM with a $1$ at a fixed position
in the top row and with connectivity $\pi$ in the FPL picture is perhaps suggestive that,
in view of trying to prove the Razumov-Stroganov conjecture,
one might rule out the existence of a {\it natural} direct bijection between FPL configurations
indexed by their link patterns
and objects counted by the entries of $\Psi_n(1)$. Indeed, if one tries to
derive some sort of action of the generators $e_i$ on each FPL configuration individually
to obtain the O(1) groundstate eigenvalue equation, what seems to come out of our
analysis is that the FPL needed to build the components of the eigenvector might not belong to the {\it same}
link pattern, but only to link patterns related via rotations. Indeed, the reshuffling of terms
when going from $\Psi_n(\pi|t)$ to $A_n(\pi|t)$, although $\Psi_n(\pi|1)=A_n(\pi|1)$,
points out to a possible mapping of various FPL configurations corresponding to
rotated versions of $\pi$.

The second conjecture of this paper relates other sums of entries of $\Psi_n(t)$ to
the corresponding sums of entries in a suitable refined TSSCPP vector $P_n(t)$, simply
built out of sums of minors of a matrix of binomial coefficients. The latter is an enumeration
of TSSCPP in the NILP picture, while keeping track both of arrival points (in bijection
with link patterns) and of the nature of the last step of each path (weight $t$ per vertical last step). 
We believe that this $t$-deformation of both TSSCPP numbers and groundstate vector
of the O(1) spin chain may be of some help for finding an explicit bijection between TSSCPP and
ASM. 

Finally, it is possible to adapt our approach to other combinatorial groundstates
of spin chains, with very analogous results. This will appear elsewhere.

\bigskip

\noindent{\bf Acknowledgments:} I am thankful to J.B. Zuber for many discussions
and his constant encouragements, and also to J. De Gier,
N. Kitanine, B. Nienhuis, P. Pearce, and P. Zinn-Justin for sharing with me their
addiction to the Razumov-Stroganov conjecture.

\vfill\eject

\appendix{A}{Refined FPL numbers}

We list below the refined FPL vectors $\alpha_n(t)=\{ \alpha_n(\pi|t)\}_{\pi\in LP_n}$, 
where $\alpha_n(\pi|t)$ as in \entryalpha.
The components of these vectors are ordered according to
the lexicographic order on link patterns, when expressed as lists of pairs of connected point labels, 
the first member of each pair being the largest of the two. For
$n=4$ for instance,
the $14$ link patterns read:
\eqn\fourteen{ \eqalign{ &\Big\{\{21,43,65,87\}, \{21,43,76,85\}, \{21,54,63,87\},\{21,54,76,83\},
\{21,65,74,83\},\cr
&\{32,41,65,87\},\{32,41,76,85\},\{32,54,61,87\},\{32,54,76,81\},\{32,65,74,81\},\cr
&\{43,52,61,87\},\{43,52,76,81\},\{43,65,72,81\},\{54,63,72,81\}\Big\}\cr}}
The component coefficients $\alpha_{n,j}(\pi)$ below are obtained by direct enumeration,
using a transfer matrix formalism for the FPL model. The vectors $\alpha_n(t)$ read, up to $n=5$:
\eqn\getphi{\eqalign{
\alpha_1(t)&=\{1\}\cr
\alpha_2(t)&=\{1,t\}\cr
\alpha_3(t)&=\{1+t,1,t,t(1+t),t^2\}\cr
\alpha_4(t)&=\{2+3t+2t^3,1+t+t^3,1+2t,2+t,1,3t^2,t^2,3t^2,t(2+5t),\cr
&t(1+2t),t^3,t(1+2t^2),t(2+t^2),t\}\cr
\alpha_5(t)&=\{7+14t+14t^3+7t^4,3+5t+6t^3+3t^4,3+6t+5t^3+3t^4,3+4t+4t^3+3t^4,\cr
&1+t+t^3+t^4,3+7t+7t^3,1+2t+3t^3,3+8t+3t^3,7+7t+3t^3,3+2t+t^3,\cr
&1+3t,3+3t,3+t,1,17t^2,6t^2,6t^2,4t^2,t^2,t^2(11+3t),t^2(3+t),\cr
&t^2(14+3t),t(7+28t+7t^2),t(3+11t+3t^2),6t^2,t(3+14t),t(3+11t),\cr
&t(1+3t),t^3(1+3t),t^4,t^3(3+3t),t(3+7t^2+7t^3),t(1+2t+3t^3),t^3(3+t),\cr
&t(3+8t+3t^2),t(7+7t^2+3t^3),t(3+2t^2+t^3),t^3,t(1+3t^2),t(3+3t^2),t(3+t^2),t\}\cr
}}

As an example, the component $\alpha_5((21)(54)(76)(98)(10\, 3)|t)=7+7t+3t^3$ 
(ninth from left in $\alpha_5(t)$) corresponds to the configurations of Fig.\seventeen.

\vfill\eject

\appendix{B} {The eigenvector $\Psi_n(t)$}

The first few Perron-Frobenius eigenvectors $\Psi_n(t)$ of $M_n(t)$ \betterx-\mono\ 
are listed below, with components 
in lexicographic order on link patterns as in appendix A. They
read, up to $n=5$:
\eqn\getpsi{\eqalign{
\Psi_1(t)&=\{1\}\cr
\Psi_2(t)&=\{1,t\}\cr
\Psi_3(t)&=\{1+t,1,t,t(1+t),t^2\}\cr
\Psi_4(t)&=\{2+3t+2t^2,1+2t,1+t+t^2,2+t,1,t(2+t),t,t(1+2t),t(2+3t+2t^2),\cr
&t(1+t+t^2),t^2,t^2(2+t),t^2(1+2t),t^3\}\cr
\Psi_5(t)&=\{7+14t+14t^2+7t^3,3+7t+7t^2,3+6t+5t^2+3t^3,3+8t+3t^2,1+3t,\cr
&3+5t+6t^2+3t^3,1+2t+3t^2,3+4t+4t^2+3t^3,7+7t+3t^2,3+3t,1+t+t^2+t^3,\cr
&3+2t+t^2,3+t,1,t(7+7t+3t^2),t(3+3t),t(3+2t+t^2),t(3+t),t,t(3+8t+3t^2),\cr
&t(1+3t),t(3+7t+7t^2),t(7+14t+14t^2+7t^3),t(3+6t+5t^2+3t^3),t(1+2t+3t^2),\cr
&t(3+5t+6t^2+3t^3),t(3+4t+4t^2+3t^3),t(1+t+t^2+t^3),t^2(3+t),t^2,\cr
&t^2(3+3t),t^2(7+7t+3t^2),t^2(3+2t+t^2),t^2(1+3t),t^2(3+8t+3t^2),\cr
&t^2(3+7t+7t^2),t^2(1+2t+3t^2),t^3,t^3(3+t),t^3(3+3t),t^3(1+3t),t^4\}\cr}}

\vfill\eject

\appendix{C}{Refined FPL numbers up to rotations}

The polynomials $\gamma_n(\pi|t)$ of eq. \defnum\ are listed below up to $n=5$.
\eqn\firfewga{\eqalign{
\gamma_1((21)|t)&=1\cr
\gamma_2((21)(43)|t)&=1+t\cr
\gamma_3((21)(43)(65)|t)&=1+2t+t^2\cr
\gamma_3((21)(54)(63)|t)&=1+t+t^2\cr
\gamma_4((21)(43)(65)(87)|t)&=2+5t+5t^2+2t^3\cr
\gamma_4((21)(43)(76)(85)|t)&=4+8t+8t^2+4t^3\cr
\gamma_4((21)(65)(74)(83)|t)&=1+t+t^2+t^3\cr
\gamma_5((21)(43)(65)(87)(10\, 9)|t)&=7+21t+28t^2+21t^3+7t^4\cr
\gamma_5((21)(43)(65)(98)(10\, 7)|t)&=16+41t+56t^2+41t^3+16t^4\cr
\gamma_5((21)(43)(76)(98)(10\, 5)|t)&=6+18t+22t^2+18t^3+6t^4\cr
\gamma_5((21)(43)(87)(96)(10\, 5)|t)&=5+10t+10t^2+10t^3+5t^4\cr
\gamma_5((21)(54)(63)(98)(10\, 7)|t)&=7+14t+18t^2+14t^3+7t^4\cr
\gamma_5((21)(76)(85)(94)(10\, 3)|t)&=1+t+t^2+t^3+t^4\cr}}

As in appendix A and B, the link patterns are labeled by the sequence of their pairs of connected labels.
The lengths of orbits are easily recovered by computing
\eqn\lorbi{ \ell(\pi)={\gamma_n(\pi|1)\over \Psi_n(\pi|1)}}

The reader can easily verify the data of eq.\firfewga, by summing the components 
of $\Psi_n(t)$ (as given by eq.\getpsi) or of $\alpha_n(t)$ (as given by eq.\getphi) over 
the orbits of the link patterns under rotation. For this, the only extra information needed
is the list of orbits of link patterns under rotation. These are summarized by writing a vector
of length $c_n$, with entries equal to
the orbit number $1,2,3,...$ of the corresponding link pattern:
\eqn\orbits{\eqalign{
&n=1:\ \ \ \{1\}\cr
&n=2:\ \ \ \{1,1\}\cr
&n=3:\ \ \ \{1,2,2,1,2\}\cr
&n=4:\ \ \ \{1,2,2,2,3,2,3,2,1,2,3,2,2,3\}\cr
&n=5:\ \ \ \{1,2,2,3,4,2,5,3,2,5,4,5,4,6,2,5,5,4,6,3,4,\cr
&\ \ \ \ \ \ \ \ \ \ \ \ \ \ \ \ 
2,1,2,5,2,3,4,4,6,5,2,5,4,3,2,5,6,4,5,4,6 \} \cr
}}

\appendix{D}{Refined TSSCPP numbers}

We list the first few refined TSSCP vectors $P_n(t)$, obtained via eqs.\recupt\
and \allow, up to $n=5$. The components are listed according to decreasing 
lexicographic order of the corresponding $(n-1)$-tuples of end-positions of the paths.
\eqn\pnoft{\eqalign{
P_1(t)&=\{1\}\cr
P_2(t)&=\{1,t\}\cr
P_3(t)&=\{1,1+t,t,t(1+t),t^2\}\cr
P_4(t)&=\{1,2+t,1+t,2+3t+t^2,1+2t+2t^2,t,t(2+t),t(1+t),\cr
&t(2+3t+t^2),t(1+2t+2t^2),t^2,t^2(2+t),t^2(1+2t),t^3\}\cr
P_5(t)&=\{1,3+t,2+t,6+5t+t^2,5+6t+3t^2,1+t,3+4t+t^2,2+3t+t^2,6+11t+6t^2+t^3,\cr
&5+11t+9t^2+3t^3,1+2t+2t^2,3+7t+8t^2+2t^3,3+8t+11t^2+6t^3,1+3t+5t^2+5t^3,\cr
&t,t(3+t),t(2+t),t(6+5t+t^2),t(5+6t+3t^2),t(1+t),t(3+4t+t^2),t(2+3t+t^2),\cr
&t(6+11t+6t^2+t^3),t(5+11t+9t^2+3t^3),t(1+2t+2t^2),t(3+7t+8t^2+2t^3),\cr
&t(3+8t+11t^2+6t^3),t(1+3t+5t^2+5t^3),t^2,t^2(3+t),t^2(2+t),t^2(6+5t+t^2),\cr
&t^2(5+6t+3t^2),t^2(1+2t),t^2(3+7t+2t^2),t^2(3+8t+6t^2),t^2(1+3t+5t^2),\cr
&t^3,t^3(3+t),t^3(3+3t),t^3(1+3t),t^4\}\cr}}
As an illustration, it is instructive to read the value of $P_3(t)$ from 
the pictures of Fig.\fplnilp, by including a weight $t$ per vertical final step.

\vfill\eject

\appendix{E}{Partial sums of entries of $\Psi_n(t)$ matching those of $P_n(t)$}

We first list the numbers $B_{n,m}$ of eq.\Bmndef, as we are to take partial sums of
entries from
$B_{n,m}$ to $c_n$ in both $\Psi_n(t)$ and $P_n(t)$, up to $n=5$.
\eqn\numbmn{ \eqalign{&B_{1,1}=1\cr
&B_{2,1}=1 \quad B_{2,2}=2\cr
&B_{3,1}=1 \quad B_{3,2}=3 \quad B_{3,3}=5\cr
&B_{4,1}=1 \quad B_{4,2}=6\quad B_{4,3}=11 \quad B_{4,4}=14\cr
&B_{5,1}=1\quad B_{5,2}=15\quad B_{5,3}=29 \quad B_{5,4}=38\quad B_{5,5}=42\cr
}}
We now list the partial sums of entries in both $P_n(t)$ and $\Psi_n(t)$
\phidef:
\eqn\probas{\eqalign{
\varphi_1(1|t)&=1\cr
\varphi_2(1|t)&=1+t\cr
\varphi_2(2|t)&=t\cr
\varphi_3(1|t)&=2+3t+2t^2\cr
\varphi_3(2|t)&=t(2+t)\cr
\varphi_3(3|t)&=t^2\cr
\varphi_4(1|t)&=7+14t+14t^2+7t^3\cr
\varphi_4(2|t)&=t(7+11t+7t^2)\cr
\varphi_4(3|t)&=t^2(4+4t)\cr
\varphi_4(4|t)&=t^3\cr
\varphi_5(1|t)&=42+105t+135t^2+105t^3+42t^4\cr
\varphi_5(2|t)&=t(42+88t+88t^2+42t^3)\cr
\varphi_5(3|t)&=t^2(25+41t+25t^2)\cr
\varphi_5(4|t)&=t^3(8+8t)\cr
\varphi_5(5|t)&=t^4\cr
}}

\listrefs

\end